\documentclass[aps,prb,superscriptaddress,showpacs,twocolumn,citeautoscript,notitlepage]{revtex4-1}

\usepackage{graphicx}
\usepackage{amsmath}
\usepackage{amssymb}
\usepackage{braket}
\usepackage{color}
\usepackage{gensymb}

\newcommand{\bea}{\begin{eqnarray*}}
\newcommand{\eea}{\end{eqnarray*}}
\newcommand{\bne}{\begin{equation*}}
\newcommand{\ede}{\end{equation*}}

\newcommand{\bnen}{\begin{equation}}
\newcommand{\eden}{\end{equation}}
\newcommand{\bean}{\begin{eqnarray}}
\newcommand{\eean}{\end{eqnarray}}
\newcommand{\bsen}{\begin{subequations}}
\newcommand{\esen}{\end{subequations}}

\newcommand{\bna}{\begin{array}}
\newcommand{\eda}{\end{array}}
\newcommand{\bnm}{\begin{enumerate}}
\newcommand{\edm}{\end{enumerate}}
\newcommand{\bni}{\begin{itemize}}
\newcommand{\edi}{\end{itemize}}

\renewcommand{\vec}[1]{\text{\boldmath{$ #1 $}}}

\begin{document}
\title{Observation of spin-orbit coupling induced  Weyl points and topologically protected Kondo effect in a two-electron double quantum dot}

\author{Zolt\'an~Scher\"ubl}
\affiliation{Department of Physics, Budapest University of Technology and Economics and MTA-BME "Momentum" Nanoelectronics Research Group, H-1111 Budapest, Budafoki \'ut 8., Hungary}
\author{Andr\'as~P\'alyi}
\affiliation{Department of Physics, Budapest University of Technology and Economics, H-1111 Budapest, Hungary}
\affiliation{Exotic Quantum Phases "Momentum" Research Group, Budapest University of Technology and Economics, 
H-1111 Budapest, Hungary}
\author{Gy\"orgy~Frank}
\affiliation{Department of Physics, Budapest University of Technology and Economics and MTA-BME "Momentum" Nanoelectronics Research Group, H-1111 Budapest, Budafoki \'ut 8., Hungary}
\author{Istv\'an~Luk\'acs}
\affiliation{Center for Energy Research, Institute of Technical Physics and Material Science, Budapest, Hungary}
\author{Gerg\H{o}~F\"ul\"op}
\affiliation{Department of Physics, Budapest University of Technology and Economics and MTA-BME "Momentum" Nanoelectronics Research Group, H-1111 Budapest, Budafoki \'ut 8., Hungary}
\author{B\'alint~F\"ul\"op}
\affiliation{Department of Physics, Budapest University of Technology and Economics and MTA-BME "Momentum" Nanoelectronics Research Group, H-1111 Budapest, Budafoki \'ut 8., Hungary}
\author{Jesper~Nyg\r{a}rd}
\affiliation{Center for Quantum Devices and Nano-Science Center, Niels Bohr Institute, University of Copenhagen, Universitetsparken 5, DK-2100 Copenhagen, Denmark}
\author{Kenji~Watanabe}
\affiliation{National Institute for Material Science, 1-1 Namiki, Tsukuba, 305-0044, Japan}
\author{Takashi~Taniguchi}
\affiliation{National Institute for Material Science, 1-1 Namiki, Tsukuba, 305-0044, Japan}
\author{Gergely~Zar\'and}
\affiliation{Exotic Quantum Phases "Momentum" Research Group, Budapest University of Technology and Economics, 
H-1111 Budapest, Hungary}
\author{Szabolcs~Csonka}
\affiliation{Department of Physics, Budapest University of Technology and Economics and MTA-BME "Momentum" Nanoelectronics Research Group, H-1111 Budapest, Budafoki \'ut 8., Hungary}
\email{palyi@mail.bme.hu}

\date{\today}
\maketitle

{\bf 
Recent years have brought an explosion of activities in the research of topological aspects of condensed-matter systems. Topologically non-trivial phases of matter are typically accompanied by protected surface states or exotic degenerate excitations such as Majorana end states\cite{Lutchyn,Oreg} or Haldane's localized spinons\cite{Haldane_nonlinear,Haldane_continuum}. Topologically protected degeneracies can, however, also appear in the bulk. An intriguing example is provided by Weyl semimetals, where topologically protected electronic band degeneracies and exotic surface states emerge even in the absence of interactions\cite{Herring,Murakami,Armitage}. Here we demonstrate experimentally and theoretically that Weyl degeneracies appear naturally in an interacting quantum dot system, for specific values of the external magnetic field. These magnetic Weyl points are robust against spin-orbit coupling unavoidably present in most quantum dot devices\cite{Nowack-esr,NadjPerge-spinorbitqubit}. Our transport experiments through an InAs double dot device placed in magnetic field reveal the presence of a pair of Weyl points, exhibiting a robust ground state degeneracy and a corresponding  protected Kondo effect.}

Mathematical tools borrowed from topology  find more and more applications in contemporary condensed-matter physics. In Weyl semimetals\cite{Turner,Armitage}, for example, the electronic band structure exhibits  isolated degeneracy points\cite{Herring}, where two bands touch. In three-dimensional systems,  these  degeneracy points can be protected by topology --- and classified by a suitably chosen Chern number: continuous perturbations may displace these  \emph{Weyl points} in momentum space, but cannot break their degeneracy. Weyl point related degeneracies of electronic states in molecules termed \emph{conical intersections} are also thought to play a fundamental role in various phenomena in photochemistry\cite{Yarkony_rmp}. They have also been predicted to appear in the context of multi-terminal Josephson junctions\cite{Riwar} and that of photonics,\cite{WenlongGao} and have also been engineered and demonstrated in coupled superconducting qubits.\cite{Schroer_chern,Roushan} 
 
\begin{figure}[b]
\begin{center}
\includegraphics[width=1\columnwidth]{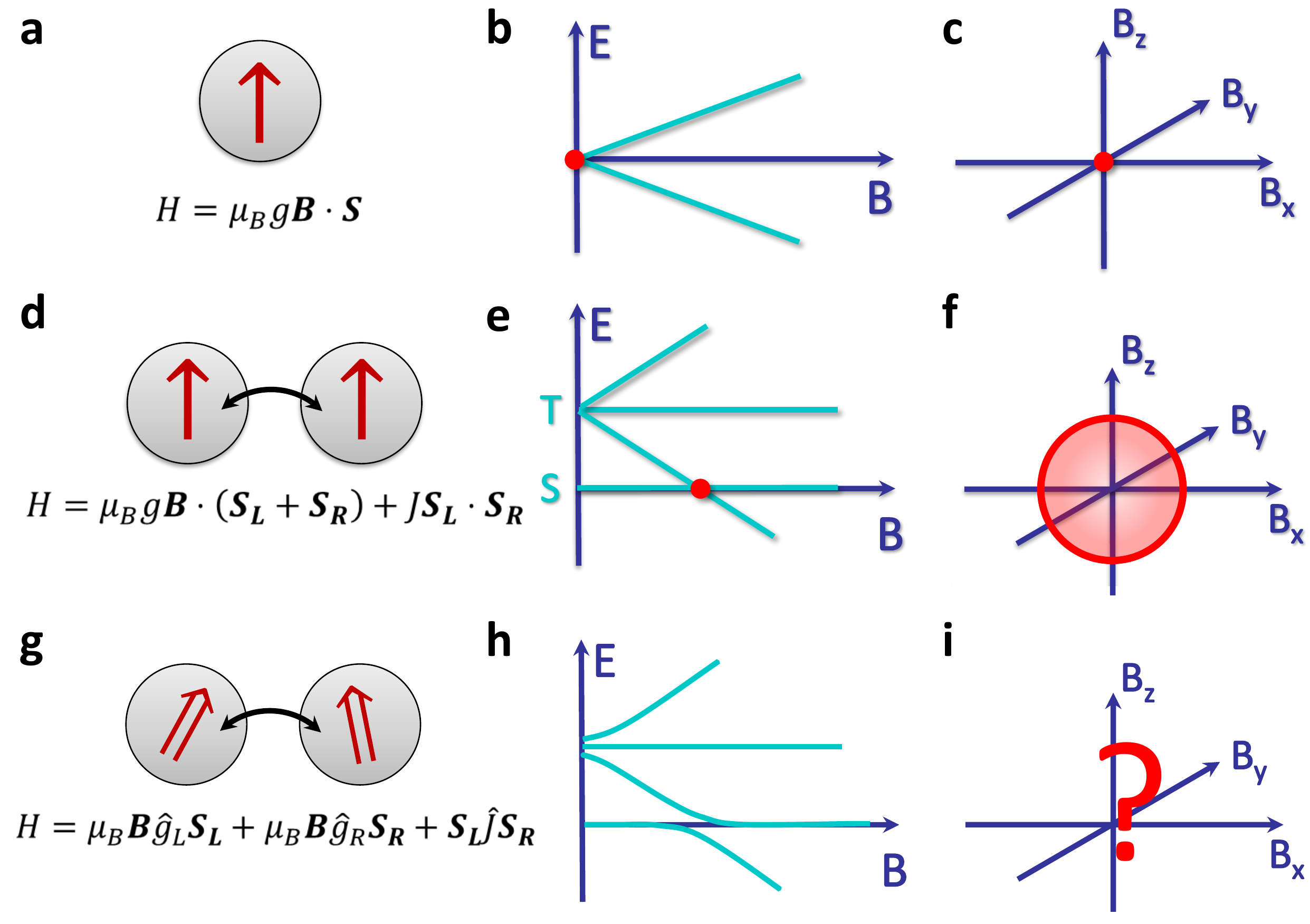}
\end{center}
\caption{{\bf Geometries of degeneracy points for simple spin systems in a Zeeman field.} {\bf a}, A single spin-1/2. {\bf d}, Two interacting $S=1/2$ spins with isotropic antiferromagnetic exchange. {\bf g}, Two interacting $S=1/2$ spins subject to spin-orbit interaction. {\bf b}, {\bf e}, {\bf h}, Characteristic magnetic-field dependence of the energy spectrum. {\bf c}, {\bf f}, {\bf i}, Geometry of  the magnetic-field values where the ground state is degenerate.
\label{fig:spintable}}
\end{figure}

The simplest example of a Weyl point arises  when a spin-1/2 electron is placed in a homogeneous magnetic field (see Fig.~\ref{fig:spintable}a-c). In this example, the parameter space is spanned by the magnetic-field vector $\vec B= (B_x,B_y,B_z)$, and the two energy eigenstates are degenerate at $\vec B = 0$. We can associate a nonzero topological charge to this degeneracy point: the ground state Chern number $C({\cal S}) = 1$ evaluated on an arbitrary closed surface ${\cal S}$ surrounding the degeneracy point (see Methods and Supplementary Information for details). This nonzero Chern number promotes this $\vec B=0$ degeneracy point to a Weyl point, and underlines the robustness of its (Kramers) degeneracy against perturbations.  

Let us now turn to the case of two coupled interacting spins, and investigate the fate of Weyl points in the presence of -- possibly strong -- spin-orbit interaction (SOI). In the most general case, this system is described by the Hamiltonian $H = H_\text{Z} + H_\text{int}$, where $H_\text{Z} = \sum_{\alpha,\beta} \mu_B B_\alpha (\hat {g}_L^{\alpha\beta} S^\beta_L + \hat {g}_R^{\alpha\beta} S^\beta_R)$ describes the Zeeman-coupling and $H_\text{int} = \sum_{\alpha,\beta} {\hat J} ^{\alpha\beta} \,S_L^\alpha S^\beta_R$ is just the exchange interaction. The SOI appears here through the anisotropic and dot dependent $g$-tensors, ${\hat {\vec g}}_{L/R}$, and the anisotropic exchange coupling tensor, ${\hat{\vec J}}$. 

In the absence of SOI (Fig.~\ref{fig:spintable}d-f), the $g$-tensors as well as the exchange coupling are just scalars, ${\hat{\vec g}}_{L/R}\to g_{L/R}$ and ${\hat{\vec J}}\to J$. The energy spectrum (Fig.~\ref{fig:spintable}e) is therefore isotropic as a function of the magnetic field. For an antiferromagnetic coupling, the ground state becomes degenerate at a sphere of radius $B =J/(\mu_B g)$, where a singlet to triplet transition occurs (Fig.~\ref{fig:spintable}f). 
 
One would expect that a small SOI would mix the singlet and triplet states close to the sphere of degeneracy, and thereby remove immediately the degeneracy. This expectation is, however, not always right. To see this consider a very large magnetic field, and calculate the corresponding Chern number, $C_\infty$. In this limit, each spin follows just the external field, yielding a Chern-number, $C_\infty = \text{sign} \det(\hat{\vec g}_L)+ \text{sign} \det(\hat{\vec g}_R)\equiv C_\infty^L + C_\infty^R$. In case of a small SOI, the $g$-tensors are close to the unit tensor, and we simply obtain $C_\infty =2$. Since, by definition, $C_\infty$ counts the total topological charge carried by the degeneracy points in the entire magnetic field space, the finiteness of $C_\infty$ signals the existence of ground state degeneracies with nonzero topological charge, typically located at single points, which we call \emph{magnetic Weyl points}. Time reversal constrains the locations of these points (see Methods): a degeneracy point at $\vec B_0$ must have  a partner at $-\vec B_0$, carrying the \emph{same} topological charge.

\begin{figure*}
\begin{center}
\includegraphics[width=2\columnwidth]{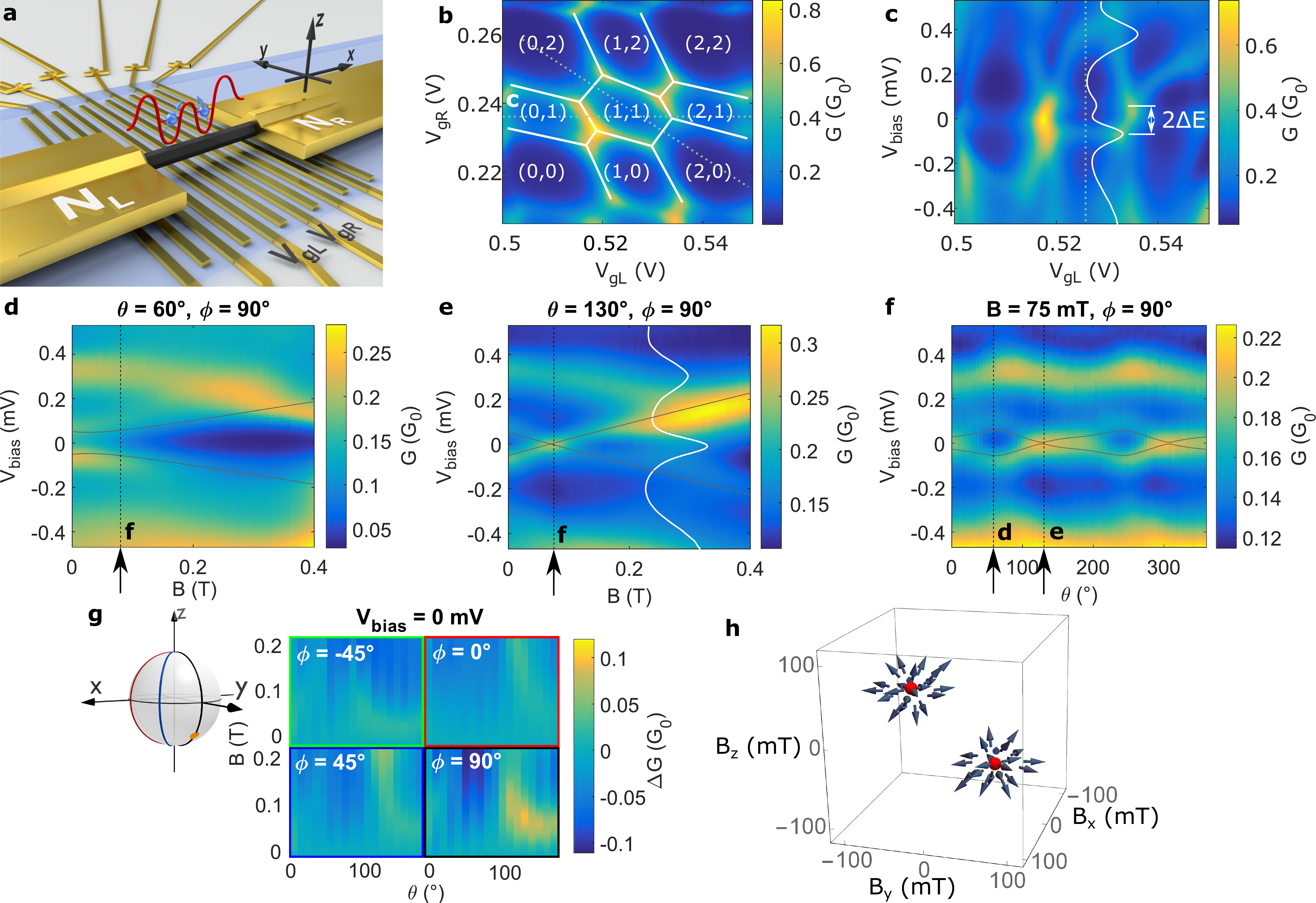}
\end{center}
\caption{{\bf Detecting magnetic Weyl points
through the conductance of a two-electron double quantum dot.}
{\bf a}, Device layout, showing the nanowire (black), and the metallic electrodes (gold) including the contacts $\text{N}_\text{L}$,  $\text{N}_\text{R}$ and the finger gates below the nanowire. The gate-controlled electric double-well potential (red) confines one electron (blue) in each well.  
{\bf b}, Charge stability diagram: zero-bias  conductance at zero magnetic field as function of two gate voltages. Labels such as (1,1) specify the number of electrons on each dot. 
{\bf c}, Finite-bias differential conductance at zero magnetic field along the dotted horizontal line in panel {\bf b} at $V_{\text{gR}}=0.236$~V, indicating an exchange splitting $\Delta E \approx J_0 \approx 0.055$~meV.
{\bf d}, {\bf e}, {\bf f}, Magnetic-field dependence of the finite-bias conductance in the (1,1) charging state.
{\bf d}, Data taken in a generic direction (here $\theta = 60\degree$ and $\phi = 90\degree$)  exhibit no ground state degeneracy. Solid gray lines are ground state  energy gaps obtained from theoretical fits (see Methods). 
{\bf e}, In the "sweet" direction, $\theta \approx 130\degree$ and $\phi \approx 90\degree$, a ground state degeneracy (a magnetic Weyl point) emerges at  $B \approx 70 \, \text{mT}$.
{\bf f}, $\theta$ dependence of the gap for $\phi=90\degree$ and a magnetic field  very close to the Weyl point, $B=75\;{\rm mT}\approx B_0$. Time reversed Weyl points emerge at $\theta \approx 130\degree$ and $\theta \approx 310\degree$.
{\bf g}, Magnetic-field and $\theta$ dependence of the zero-bias magnetoconductance, $\Delta G(B) = G(B)-G(B=0)$ along the lines indicated on the left sketch by colored lines. The maximum at $B\approx 70 {\rm mT}$ in the bottom right panel indicates a magnetic Weyl point, also marked on the surface of the sphere.
{\bf h}, Visualization of the calculated ground state  Berry curvature vector field in the vicinity of the two magnetic Weyl points (red). The outward oriented hedgehog patterns indicate that the two Weyl points carry the same topological charge.}
\label{fig:doubledot}
\end{figure*}

By these topological considerations, we expect that in case of $C_\infty^L= C_\infty^R = 1$, two Weyl points at $\pm \vec B_0$ carry the total topological charge $C_\infty = 2$ (red points in Fig.~\ref{fig:doubledot}h). Using random spin Hamiltonians as well as random two-site Hubbard models (see Methods and Supplementary Information), we have numerically verified  that this scenario of two magnetic Weyl points is generic, and is indeed realized in over 99\% of randomly generated two-spin Hamiltonians. These magnetic Weyl points are topologically `robust' in the sense that although they can move around in the magnetic parameter space upon continuous deformations of the Hamiltonian, they cannot suddenly disappear, and the spectrum remains degenerate in them. 

To demonstrate experimentally the existence of magnetic Weyl points in a spin-orbit-coupled interacting two-spin system, we carried out low-temperature  electric transport measurements through a serial  InAs nanowire double quantum dot (DQD) \cite{FasthPRL2007,SchroerPRL2011,PfundPRL2007,FasthNanoLett2005,WangArXiv2018} in the temperature range 60-300 mK. The setup is sketched in Fig.~\ref{fig:doubledot}a. (For sample fabrication and characterization see Methods.) Alternative experimental techniques to explore these magnetic Weyl points are Landau-Zener \cite{Reilly,Petta_beamsplitter,Fogarty} or EDSR spectroscopy \cite{NadjPerge}, as applied to various two-electron double-dot devices.

In the experiments we focused on the (1,1) charge configuration of the device (see Fig.~\ref{fig:doubledot}b), where the DQD contains two spatially separated and exchange-coupled spins. In this region, we expect  that the ground state of the system is a singlet, and the first excited state separated by $\Delta E\approx J_0$ is a triplet (see Supplementary Information). The finite exchange splitting $J_0 \approx 0.055\, \text{meV}$ is demonstrated by the bias-dependent differential conductance data presented in Fig.~\ref{fig:doubledot}c. At the center of the (1,1) configuration, that is, along the vertical dashed line, the conductance is suppressed at small biases, but increases once the bias is sufficiently high to induce inelastic co-tunneling processes populating the triplet states. The differential conductance $G = dI/dV_\text{bias}$ (white curve) has therefore two finite-bias peaks (white lines) placed symmetrically at the first excited state of the DQD, at $e V_\text{bias} = \pm \Delta E\approx \pm J_0$. The asymmetry $G(V_\text{bias}) \neq G(-V_\text{bias})$ can be attributed to asymmetric coupling to the leads. 

We now switch on the magnetic field to tune the relative energies of the ground and excited states, and explore by the co-tunneling spectroscopy outlined above, how the energy gap $\Delta E = \Delta E(\vec B)$ between the ground and first excited states varies with the field (Fig.~\ref{fig:spintable}h).\cite{Jeong,Chorley,Spinelli} Two examples are shown in Figs.~\ref{fig:doubledot}d and e, where we present the conductance $G(B,V_\text{bias})$ for magnetic fields $\vec B = B (\sin\theta \cos \phi,\sin\theta \sin \phi, \cos \theta)$ oriented along two different directions (see reference frame in Fig.~\ref{fig:doubledot}a).

In Figs.~\ref{fig:doubledot}d and e, the magnetic field dependence of the gap $\Delta E(\vec B)$ is traced by the large-conductance features close to zero bias, also indicated by solid lines. The observed behavior is markedly different in the two cases: Fig.~\ref{fig:doubledot}d displays a behavior in line with the naive mixing argument, and the gap remains open for all values of $B$. In Fig.~\ref{fig:doubledot}e, however,  the gap closes at around $B_0=70 \, \text{mT}$, where a zero-bias conductance peak develops (white continuous line), suggesting that this magnetic field vector corresponds to  a magnetic Weyl point. 

The scenario of the two magnetic Weyl points at opposite  magnetic fields $\pm \vec{B}_0$, fits perfectly our experimental observations. To demonstrate this, we display the conductance $G(\theta,V_\text{bias})$ in Fig.~\ref{fig:doubledot}f for a fixed magnetic-field length $B=75 \, \text{mT}$ and $\phi = 90^\circ$, while varying  the polar angle $\theta$ over a range of 360 degrees. Our data indicate ground state degeneracies at two opposite isolated points, $\theta \approx 130^\circ$ and $\theta \approx 310^\circ$, but a finite gap otherwise. The solid lines in  Figs.~\ref{fig:doubledot}d,e,f, indicating the gap, are not only guides to the eye: they were computed from a two-site Hubbard model (see Methods), with parameters adjusted to yield a good overall match to experimental observations. Fig.~\ref{fig:doubledot}h visualizes the Berry curvature fields (see Supplementary Information) and the associated topological charges at the two Weyl points, as computed  numerically from this two-site Hubbard model. 
 
We  support further the  scenario of the two magnetic Weyl points by showing a more complete scan of the zero-bias magnetoconductance $\Delta G(B) = G(B) - G(B=0)$ in Fig.~\ref{fig:doubledot}g. The four panels of Fig.~\ref{fig:doubledot}g correspond to four azimuthal angles, $\phi \in \{-45^\circ, 0^\circ, 45^\circ, 90^\circ\}$ of the magnetic field, as depicted in the sketch on the left side of panel g. Each panel of Fig.~\ref{fig:doubledot}g displays the zero-bias magnetoconductance  $\Delta G(\theta,B)$ as the function of the polar angle $\theta$ and strength $B$ of the magnetic field. The most prominent local maximum in the bottom right panel of Fig.~\ref{fig:doubledot}g indicates that the two-electron double dot  has a magnetic Weyl point close to that region, $\phi\approx 90\degree$, $\theta\approx 130\degree$ and $B\approx 70 \;{\rm mT}$ (also seen in Fig.~\ref{fig:doubledot}e). 

\begin{figure}
\begin{center}
\includegraphics[width=1\columnwidth]{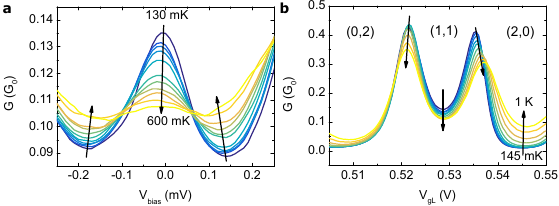}
\end{center}
\caption{{\bf Protected two-electron Kondo effect in a double quantum dot.}
{\bf a}, Temperature dependent conductance in the center of the (1,1) charge configuration (corresponding to $V_\text{gL} = 0.523$ V, $V_\text{gR} = 0.236$ V), at the magnetic Weyl point ($B = 60$ mT, $\theta = 130^\circ$, $\phi = 90^\circ$). {\bf b}, Temperature- and gate-voltage dependence  of the zero-bias conductance along the diagonal dashed line in Fig.~\ref{fig:doubledot}b. Magnetic field as specified above.}
\label{fig:kondo}
\end{figure}

In our device, the topologically protected degeneracy is accompanied by an increased zero-bias conductance in the vicinity of the magnetic Weyl points (see white curve on Fig.~2e). This increased conductance is due to a two-electron Kondo effect,\cite{Jeong,Chorley,Spinelli} as clearly revealed by the temperature and voltage dependence of our transport data in Fig.~\ref{fig:kondo}, complying with the Kondo behavior seen in other experiments\cite{GoldhaberGordon,vanderWiel_kondounitary,Kouwenhoven}. The differential conductance at the Weyl point exhibits, in particular, a pronounced zero-bias Kondo peak with a height increasing upon decreasing temperature (see Fig.~\ref{fig:kondo}a). This increased low temperature conductance appears to be  characteristic of the whole charge (1,1) domain, as demonstrated in  Fig.~\ref{fig:kondo}b, presenting the temperature dependence of the zero-bias conductance along the diagonal dashed line in Fig.~\ref{fig:doubledot}b. In contrast, in the regions corresponding to (2,0) and (0,2) charge configurations, the ground state is unique; there the conductance shows thermal activation, and is suppressed  with decreasing temperatures.  

So far, we have argued that for two  interacting spins the appearance of ground state degeneracies at a pair of time-reversed magnetic Weyl points is
natural and generic. One can, however, engineer Hamiltonians exhibiting very different degeneracy structures, still consistent with our topological arguments: 
(i) In the presence of an antiferromagnetic exchange coupling with SU(2) spin symmetry (no SOI) degeneracy points appear on a sphere, as shown in Fig.~\ref{fig:spintable}f.
(ii) A single degeneracy point of topological charge +2 at $\vec B=0$ appears for an isotropic ferromagnetic interaction. 
(iii) An even number $N_-$ of Weyl points of topological charge $-1$, together with $N_+ = N_- +2$ of Weyl points with topological charge $+1$ can appear. The case $N_- = 2$ and $N_+ = 4$ is realized, e.g., when the exchange interaction is isotropic and antiferromagnetic, and the principal directions of the two $g$-tensors are aligned, but the six principal values are all different.  
(iv) A strong  SOI can also change the sign of the determinant of a $g$-tensor. For ${\rm det}\{{\hat{\vec g}}_{L}\}{\rm det}\{{\hat{\vec g}}_{R}\}<0$, e.g.,  the topological charge of the two spins cancels, $C_\infty = 0$, and magnetic Weyl points and degeneracies may be completely absent in any field. Note that the sign of the determinant of a $g$-tensor is irrelevant as long as we consider the Zeeman splitting of a single spin. However, it gains immediate physical relevance once multiple spins are considered. 

We remark that only cases (iii) and (iv) above are topologically stable and robust under small changes in the Hamiltonian. All other cases turn out to be fragile in the sense that a fine-tuning of the model parameters is needed in the presence of SOI to realize them (see Supplementary Information).

Remarkably, the argument applied to two coupled spins can be generalized to interacting \emph{multi-spin} systems. In fact, the asymptotic ground state Chern number for $N$ non-interacting spins with size $1/2$ and isotropic $g$-tensors\cite{Gritsev} is $C_\infty = N$. This suggests that for a generic $N$-spin system there are $N$ magnetic field values where Weyl points of topological charge +1 appear, and the ground state is degenerate. For even values of $N$, these degeneracies must appear as time reversed pairs, while for an odd number of spins a Weyl point must appear at $\vec B = 0$, as also implied by Kramers' theorem. These arguments can be readily extended to systems of spin $S$  impurities, too, where the total topological charge adds up to $C_\infty = 2S\;N$. 

We thus establish that magnetic Weyl degeneracies are generic in interacting electron systems in the presence of SOI. The magnetic Weyl points demonstrated here and the topologically protected ground state degeneracies are analogous to the degeneracy points in electronic band structures of Weyl semimetals, and have important physical implications such as the corresponding topologically protected Kondo effect observed.

{\bf Acknowledgments}

We acknowledge  J.~Asb\'oth, \'A.~Butykai, W.~Coish, K.~Grove-Rasmussen, B.~Hensen, R.~Maurand, P.~Nagy, J.~Paaske, F.~Pollmann, and T.~Tanttu are for valuable discussions, and  M.~H.~Madsen and C.~B.~S{\o}rensen for technical support.
This work was supported by the National Research Development and Innovation Office of Hungary within the Quantum Technology National Excellence Program (Project No. 2017-1.2.1-NKP-2017-00001), under OTKA Grant 124723, 127900, by the New National Excellence Program of the Ministry of Human Capacities,  by CA16218 nanocohybri COST Action, and by the Danish National Research Foundation, and by the Elemental Strategy Initiative conducted by the MEXT, Japan and the CREST (JPMJCR15F3), JST. 

{\bf Author contributions:}
Experiments were designed by S.~C., wires were developed by J.~N., hBN crystals by K.~W. and T.~T., devices were developed and fabricated by Gy.~F., I.~L., B.~F., G.~F. and Sz.~Cs., measurements were carried out and analyzed by Z.~S., Gy.~F. and S.~C. Theoretical interpretation was given by A.~P., Z.~S., Gy.~F. and G.~Z.. A.~P., Z.~S. and G.~Z. prepared the manuscript. 

\vskip1cm
{\bf \noindent \large Methods:}

\vskip0.5cm
{\bf\noindent Sample fabrication and measurement details} \vskip0.2cm 
An array of Cr/Au (with 5/25 nm thickness)  bottom gates (see Fig.~\ref{fig:doubledot}a) with a width of  40 nm  and a period of 100 nm  was prepared by e-beam lithography and e-beam evaporation on a Si/SiO$_2$ substrate. Exfoliated hexagonal boron-nitride (hBN) flakes with a thickness of 20~nm were positioned on top of the bottom gates by a transfer microscope to electrically isolate the bottom gate electrodes from the nanowire. The  80~nm diameter InAs nanowire   was placed on the hBN by a micromanipulator setup. The nanowire and the bottom gates were contacted by Ti/Au electrodes (10/80 nm), defined in a second e-beam lithography and e-beam evaporation step.

The sample was measured in Leiden Cryogenics CF-400 cryo-free dilution refrigerator, equipped with a two-dimensional vector magnet. To vary the magnetic field in three dimensions, the sample holder probe was rotated manually to 4 different  orientations $\phi \in \{-45^\circ, 0^\circ, 45^\circ, 90^\circ\}$. After each rotation, the base temperature was different due to the different thermal contact between the probe and the cryostat. The differential conductance of the DQD was measured in a two-point geometry by lock-in technique at 237 Hz with 10 $\mu$V ac excitation with a home-built $I/V$ converter. The conduction band was not fully depleted by the gates:  charge-configuration labels in Fig.~\ref{fig:doubledot}b therefore correspond to the number of electrons above closed shells in each quantum dot holding an unknown, large number of electron pairs.

\vskip0.5cm
{\bf\noindent Berry curvature and Chern number}  \vskip0.2cm
Consider the ground state manifold $\psi_0(\vec B)$ of a family of Hamiltonians $H(\vec B)$ parametrized by the magnetic field, $\vec B$. Assuming that $\psi_0$ is differentiable in the vicinity of $\vec B$, we define the Berry connection vector field $\mathcal{A}=(\mathcal{A}_x,\mathcal{A}_y,\mathcal{A}_z)$ as
\bean \mathcal{A}(\vec B) \equiv i \braket{\psi_0(\vec B) | \nabla_\vec{B} | \psi_0(\vec B)}. \eean  
The Berry curvature vector field $\mathcal B=(\mathcal{B}_x,\mathcal{B}_y,\mathcal{B}_z) $ is defined as the curl of the Berry connection, 
\bean \mathcal{B}(\vec B) = \nabla_\vec{B} \times \mathcal{A}(\vec B). \eean
Notice that while the Berry connection $\mathcal{A}$ is gauge dependent, the Berry curvature $\mathcal{B}$ is not.

Consider now a closed surface $\cal S$ in the magnetic-field space, such that the ground state is non-degenerate at any point of $\cal S$. The (ground state) Chern number associated with this surface is then
\bean C({\cal S}) = \frac 1 {2\pi} \oint_{\cal S} d \vec s \cdot \mathcal{B}. \eean
For details, see Sec.~IV of Supplementary Information.

\vskip0.5cm
{\bf\noindent Magnetic Weyl points form time-reversed pairs} \vskip0.2cm

If there is a magnetic Weyl point at $\vec B_0$,  then -- by time reversal -- there is also one at $-\vec B_0$.  
This follows from the properties of time reversal, $\tau$. 
(i) $\tau$ is an antiunitary operator, that is $\braket{\tau \varphi |\tau \psi} = \braket{\varphi|\psi}^*$ for any $\varphi$ and $\psi$.
(ii) 
$\tau$ changes the sign of each spin operator,  hence $\tau H(\vec B) \tau^{\dagger} = H(-\vec B)$. (iii) From (ii) it follows that   if $H(\vec B) |\psi \rangle = E |\psi \rangle $, then $H(-\vec B) \tau |\psi\rangle = E\, \tau  |\psi\rangle$. Thus, apart from an overall phase, $ \tau |\psi(\vec B) \rangle =  |\psi(-\vec B)\rangle$. Thus a degeneracy at $\vec B_0$ implies a degeneracy at $-\vec B_0$, and at non-degenerate points $\mathcal{B}(\vec B) = -\mathcal{B}(-\vec B)$. 

\vskip0.5cm
{\bf\noindent Two-site Hubbard model of the double quantum dot} 
\vskip0.2cm

Theoretical results in Figs.~\ref{fig:doubledot}d, e, f, h were produced by a spin-orbit coupled two site Hubbard model, with Hamiltonian $H = H_0 + H_\text{Z}$. Here, the Hamiltonian in the absence of magnetic field is
\begin{eqnarray}
H_0 &=&\frac{U_L}{2} n_L (n_L-1)  + \frac{U_R}{2} n_R (n_R -1) + \varepsilon_L n_L +\varepsilon_R  n_R \nonumber \\
&+& \sum_{ss' \in \{\pm\}} \left(t^{ss'} c^\dag_{Ls} c_{Rs'} + h.c. \right),
\end{eqnarray}
with $U_{L/R}$ the strength of the Coulomb interaction on the left/right dot, $n_{L/R}$  the occupation numbers, $\varepsilon_{L/R}$ the gate-controlled on-site energies, and $t^{ss'} = t_0 \delta_{ss'} -i \sum_{\alpha=(x,y,z)} t_{\alpha} \sigma^\alpha_{ss'}$ a spin-flip hopping term\cite{Danon}, with real-valued hopping amplitudes $t_0, t_x, t_y, t_z$. The $\sigma^\alpha$ here denote Pauli matrices, and  are related to the spin operators in the usual way, e.g., $S_L^x = \sum_{ss'}c^\dag_{Ls}\sigma^{x}_{ss'} c_{Ls'}$. In an external magnetic field, we also add the Zeeman terms $H_\text{Z} = \mu_B \vec B \cdot \left( \hat{\vec g}_L \vec S_L+ \hat{\vec g}_R \vec S_R \right)$. The spin dependent interdot hopping as well as the nontrivial $g$-tensors can be attributed to strong spin-orbit interaction in the InAs nanowires\cite{FasthPRL2007,SchroerPRL2011,PfundPRL2007,WangArXiv2018}.

We have determined the values of the model parameters to provide a good overall agreement with the experimentally observed. For the methodology, see the Supplementary Information. These parameters were then used to derive the theoretical results in Figs.~\ref{fig:doubledot}d, e, f, h. The $g$-tensors used were
\bean
\label{eq:gtens} 
\hat{\vec g}_{L} &=& \begin{pmatrix} 2.136 & -1.089 & 0.443 \\ -1.089 & 11.696 & -5.315 \\ 0.443 & -5.315 & 6.617 \end{pmatrix}, \\
\hat{\vec g}_{R} &=& \begin{pmatrix} 8.739 & -1.703 & 5.835 \\ -1.703 & 8.637 & 1.532 \\ 5.835 & 1.532 & 14.713 \end{pmatrix}.
\eean
Hoppings were set to $t_0=0.0525~\mathrm{meV}$, $t_x=-0.0151~\mathrm{meV}$, $ t_y=0.0565~\mathrm{meV}$, $t_z=-0.0697~\mathrm{meV}$, and Coulomb energies to $U_L = 1$ meV, $U_R = 0.6$ meV. 
The on-site energies corresponding to the center of the (1,1) hexagon of the charge stability diagram in Fig.~\ref{fig:doubledot}b read $\varepsilon_L = -U_L/2$ and $\varepsilon_R = -U_R/2$.

\bibliography{weyl}

\end{document}